\documentclass[aps,prd,nofootinbib]{revtex4}
\usepackage[colorlinks=true, pdfstartview=FitV, linkcolor=blue, citecolor=red, urlcolor=magenta]{hyperref}
\usepackage{graphicx}
\usepackage{latexsym}
\usepackage{amsmath}
\usepackage{amsfonts}
\usepackage{amssymb}
\usepackage{verbatim}
\usepackage{braket}
\usepackage{extarrows}

\newcommand{\be}{\begin{equation}}
\newcommand{\ee}{\end{equation}}
\newcommand{\bea}{\begin{eqnarray}}
\newcommand{\eea}{\end{eqnarray}}

\newcommand{\ben}{\begin{eqnarray}}
\newcommand{\een}{\end{eqnarray}}

\begin{document}

\title{Quantum vacuum fluctuation effects in a quasi-periodically identified conical spacetime}


\author{$^{1}$Klecio E. L. de Farias}
\email{klecio.limaf@gmail.com}

\author{$^{1}$ Herondy F. Santana Mota}
\email{hmota@fisica.ufpb.br}

\affiliation{$^{1}$Departamento de F\' isica, Universidade Federal da Para\' iba,\\  Caixa Postal 5008, Jo\~ ao Pessoa, Para\' iba, Brazil.}


\begin{abstract}
We consider a quasi-periodically identified conical spacetime, like the one of a cosmic string or disclination, to investigate nonzero averaged quantum vacuum fluctuations effects on the energy-momentum tensor and induced current density associated with a charged scalar field. We obtain exactly closed analytical expressions for the two-point Wightman function and the vacuum expectation value of the field squared, as well as for all components of the energy-momentum tensor. As to the induced current density, due to the quasi-periodic condition used, the only nonzero averaged component found is the azimuthal one. We also compare our results with previous ones found in literature.
\end{abstract}
 \maketitle


\section{Introduction}
\label{intro}
Quantum vacuum fluctuation effects arise every time the quantum vacuum of a relativistic field (scalar, vector, fermionic) is somehow perturbed, either by the implementation of boundary conditions of some sort or by external fields \cite{Milonni:1994xx, birrell1984quantum, bordag2009advances, Milton:2001yy}. These effects also take place in curved spacetimes, in principle, solely as a consequence of the existence of a nontrivial topology  \cite{birrell1984quantum, bordag2009advances, Milton:2001yy, Mota:2017slg, saharian2010topological}. Evidently, boundary conditions or/and external fields can also be considered in these spacetimes with nontrivial topology, in which case an effect associated with each one of these ingredients, separately, comes about. One could mention, as examples of quantum vacuum effects, fluctuations of the classical light-cone \cite{Ford:1994cr, Mota:2016mhe}, induced Brownian motion \cite{Ford:2005rs, Bessa:2008pr, Gour:1998my}, induced current density of a charged field \cite{deMello:2014ksa, Braganca:2014qma, Santos:2018ttf}, self-interactions \cite{BezerradeMello:2012nq, GL, PhysRevD.65.085013} and the mostly known Lamb shift  \cite{Milonni:1994xx}, Casimir effect \cite{Milonni:1994xx, birrell1984quantum, bordag2009advances, Milton:2001yy,  Mostepanenko:1997sw} and several others. Here we will focus attention on the energy-momentum tensor and current density as nonzero averaged physical observables, in particular, the 00-component of the energy-momentum tensor which is the vacuum energy density arising from a Casimir-like effect.

As originally envisaged, and already experimentally verified \cite{Sparnaay:1958wg, Lamoreaux:1996wh, PhysRevLett.81.5475, mohideen1998precision, Bressi:2002fr, MOSTEPANENKO2000}, the Casimir effect arises when two discharged parallel plates are placed together in vacuum, very closely, at a distance of the order of $\mu m$, resulting in an attractive interaction between the plates, as a consequence of the modification of the quantum vacuum fluctuations of the electromagnetic field, the dominant fluctuations over the other fields \cite{Casimir:1948dh}. The calculation of the intensity of the attractive force, of course, is performed by considering the Minkowski spacetime. However, theoretically, many works have been done considering the Casimir effect in spacetime with nontrivial topology, along with boundary conditions \cite{birrell1984quantum, bordag2009advances, Milton:2001yy, Mostepanenko:1997sw}. In this context, other components of the energy-momentum tensor have also been considered \cite{Mota:2017slg, saharian2010topological, BezerradeMello:2011nv, BezerradeMello:2011sm, Christensen:1976vb, Wald:1978pj, smith, Braganca:2019mvj, Santos:2019ais, Sriramkumar:2000nq}.

Another interesting physical observable that is worth investigating is the induced current density associated with a charged field. In the context of the semi-classical approach of quantum field theory, for instance, the vacuum expectation value of the current density can be used as source in the Maxwell equations providing a further understanding of the electromagnetic field dynamics. This line of research have been considered in several works in the past few years \cite{deMello:2014ksa, Braganca:2014qma, Santos:2018ttf, Sriramkumar:2000nq} and our intension is also to follow up in this direction. Note that, in the absence of a gauge field, there can exist an induced current density when, for instance, a quasi-periodic boundary condition is adopted, which is the case we will be considering.

A very interesting spacetime with nontrivial topology is the one generated by the gravitational field of a idealized thin, straight and very long cosmic string \cite{VS, hindmarsh, escidoc:153364}. Cosmic string is a line-like topological defect predicted in many extensions of the Standard Model of particle physics, and more recently in the context of string theory \cite{Copeland:2011dx, Hindmarsh:2011qj}. It can have several cosmological, gravitational and astrophysical consequences \cite{VS, hindmarsh, escidoc:153364, Copeland:2011dx, Hindmarsh:2011qj, Mota:2014uka}, which makes its physics of great interest. It should be pointed out that the counterpart of a cosmic string in condensate matter is a disclination \cite{Katanaev:1992kh}. Both, cosmic string and disclination, are described by a conical spacetime.

The Casimir effect can be used to understand the behaviour of materials in nanoscale and, in this sense, also may appear once one considers, for instance, tube of carbon atoms at a scale of $10^{-9}m$ (nanotubes), the exactly scale where the Casimir effect makes itself stronger. An analogue model may well be pictured in which a quasi-periodic boundary condition mimics properties of nanotubes. The quasi-periodic boundary condition is introduced by means of the identification $\Phi(t,r,\varphi,z)=e^{2\pi i\beta}\Phi(t,r,\varphi+2\pi/q,z)$ \cite{saharian2010topological}, causing an explicit dependence on the parameter $\beta$ by the solution of the scalar field obeying the condition. The link between the quasi-periodic boundary condition and the nanotubes physics happens through the way the carbon sheet is cut, which determines the properties of the nanotube. This is directly related to the phase angle whose values take place in the interval  $0\leq\beta\leq1$, defining the properties of the nanotube. According to the theory, the properties of a nanotube is determined by the chiral vector (also known as circumferential vector) $\vec{c}=n_1\vec{a}_1+n_2\vec{a}_2$, where $n_1$ and $n_2$ are integer numbers known as chiral indices. As the graphene present a hexagonal symmetry, the chiral angle can assume values between $0^{\circ}$and $60^{\circ}$, although the nanotube can often be found in an interval $0^{\circ}\leq2\pi\beta\leq30^{\circ}$. In this case another equivalent nanotube can also be found in the interval $30^{\circ}\leq2\pi\beta\leq60^{\circ}$ \cite{o2006carbon}. The difference lies in the helicity of the graphene lattice, where the points around the tube changes from the right side to the left one, modifying optical properties. In terms of chiral pairs $(n_1,n_2)$ there are two special cases. If we choose $\beta=0$, it characterizes a Zigzag nanotube with chiral pair $(n_1,0)$ that may represent a semimetal when $ n_1-n_2=3p$ or a semiconductor when $n_1-n_2\neq3p$, with $p\in\mathbb{Z}$. For $\beta=1/3$ the phase angle becomes $60^{\circ}$ and represents the Arm-chair nanotube, which is always a metal nanotube described by the chiral pair $(n,n)$, that is, $n_1=n_2$ . The chiral nanotube, in addtion, occurs for $0^{\circ}<2\pi\beta<30^{\circ}$ and can be semimetal or semiconductor, following the same conditions as in the Zigzag nanotube case.

The role of the scalar field in graphene theory can be interpreted as the scalar potential $\Phi(x)$ related to the free static electric field $E(x)$, given by $E(x)=-\nabla_x\Phi(x)$ \cite{malic2013graphene}. It can be used in quantum field theory to represent excitations of spinless Dirac fermions in graphene, and its fluctuations can be related to a quantum field of a plasmon in graphene \cite{phan2017functional}. Another application is in the context of graphene nanorings. In this case, depending on the form of the nanoring (armchair or zigzag) it can be characterized by a  position-dependent scalar field denoted by $m(x)$, representing a position-dependent mass \cite{romanovsky2013topological}.

The paper is organized as follows. In Sec.\ref{secII} the solution of the Klein-Gordon equation in a quasi-periodically identified conical spacetime (cosmic string or disclination) is obtained, along with a closed and exact analytical expression for the Wightman two-point function. In Sec.\ref{secIII} we explicitly calculate the vacuum expectation value of the field squared, of the energy-momentum tensor and the induced current density. Finally, in Sec.\ref{secIV} we present our conclusions. Throughout the paper we use natural units $G=\hbar=c=1$.

\section{Wightman Function}
\label{secII}
We want to investigate the changes in the quantum vacuum fluctuations of a charged scalar field arising as a consequence of a quasi-periodically identified spacetime. Generically, the spacetime to be considered here is  either the cosmic string or the disclination one, both characterized as having a conical topology. As a topological defect, their one-dimensional nature allows us to use cylindrical coordinates to write the following line element describing a $(3+1)$-dimensional conical spacetime:
\begin{equation}
ds^2=g_{\mu\nu}dx^{\mu}dx^{\nu}=dt^2-dr^2-r^2d\varphi^2-dz^2,
\label{eq0}
\end{equation}
where the cylindrical coordinates take values in the range $r\geq0$, $0\leq\varphi\leq 2\pi/q$ and $-\infty <(t,z)<+\infty$. The presence of the cosmic string is characterized by the parameter $q\geq 1$, which is given in terms of the linear mass density of the cosmic string, $\mu$, that is, $q^{-1} = 1-4G\mu$, where $G$ is the Newton's gravitational constant \cite{VS, hindmarsh, escidoc:153364}. Moreover, in condensate matter systems, the parameter $q$ characterizes a disclination and can take on any values such that $q\geq 0$ \cite{Katanaev:1992kh}.

The scalar field modes propagating in the conical spacetime \eqref{eq0} is to obey a quasi-periodic boundary condition given by
\begin{equation}
\Phi(t,r,\varphi,z)=e^{i2\pi\beta}\Phi(t,r,\varphi+2\pi/q,z),
\label{eq1}
\end{equation}
where $\beta$ is a parameter defined in the interval $0\leq\beta\leq1$ and regulates the phase angle present in \eqref{eq1}.

The charged scalar field theory we want to consider here is obtained, in (3+1)-dimensional curved spacetime, through the following action:
\begin{equation}
S=\frac{1}{2}\int dx^4\sqrt{|g|}\left[g^{\mu\nu}\nabla_{\mu}\Phi\nabla_{\nu}\Phi^{*} - (m^2+\xi R)|\Phi|^2\right],
\label{action}
\end{equation}
where $g=\text{det}(g_{\mu\nu})$, $R$ is the Ricci scalar, $\nabla_{\nu}$ is the covariant derivative and $\xi$ is the constant coupling to gravity parameter. In the massless case, the action and the scalar field obey a conformal symmetry if we choose $\xi=\frac{1}{6}$, leading to a nonminimally coupled scalar field theory invariant under conformal transformation \cite{birrell1984quantum}. Note that we are not considering a gauge field, $A_{\mu}(x)$, which is normally taken into consideration in several cases \cite{deMello:2014ksa, Braganca:2014qma, Santos:2018ttf}.

The Klein-Gordon equation for a nonminimally coupled charged scalar field $\Phi(x)$, thus, follows from the action \eqref{action} and is written as
\begin{equation}
\left[\frac{1}{\sqrt{|g|}}\partial (\sqrt{|g|}g^{\mu\nu}\partial_{\nu})+m^2+\xi R\right]\Phi(x)=0.
\label{eq1.1}
\end{equation}
Note that we are considering a conical defect that has associated to it a Dirac delta-type Ricci scalar function and, therefore, has an infinity curvature at origin but zero one at $r\neq 0$ \cite{VS, hindmarsh}. In order to avoid the infinity curvature contribution at origin, and the complications following from that (see for instance \cite{Spinally:2000ii}), we will be interested only in the minimally coupled scalar field case, $\xi=0$.

In the conical spacetime given by the line element \eqref{eq0}, the Klein-Gordon equation \eqref{eq1.1} for the minimally coupled scalar field, $\Phi(x)$,  takes the form
\begin{equation}
\left[\frac{\partial^2}{\partial t^2}-\frac{1}{r}\frac{\partial}{\partial r}\left(r\frac{\partial}{\partial r}\right)-\frac{1}{r^2}\frac{\partial^2}{\partial\varphi^2}-\frac{\partial^2}{\partial z^2}+m^2\right]\Phi(x)=0.
\label{eq1.2}
\end{equation}

Thereby, the regular solution at origin of the above equation for the minimally coupled scalar field obeying the quasi-periodic boundary condition \eqref{eq1} is given by
\begin{equation}
\Phi(t,r,\varphi,z)=Ae^{-i\omega_k t}e^{i\nu z}e^{iq(n-\beta)\varphi}J_{q|n-\beta|}(\eta r),
\label{eq2}
\end{equation}
where  $J_{\alpha}(x)$ is the Bessel function of the first kind, $A$ is a normalization constant, $\omega_k = m^2 + \eta^2 + \nu^2$ and $k=(n,\eta, \nu)$ is the set of quantum numbers.

The normalization constant, $A$, present in the solution \eqref{eq2} is calculated by using the condition
\begin{equation}
\int d^3x\sqrt{|g|}\Phi_{k}(x)\Phi_{k^{\prime}}^*(x)=\frac{1}{2\omega_k}\delta_{kk^{\prime}},
\label{eq3}
\end{equation}
where the notation $\delta_{kk^{\prime}}$ stands for the Kronecker delta in the case of the discrete quantum number $n$, and Dirac delta in the case of the continuous quantum numbers $\eta$ and $\nu$. Thus, using Eqs. \eqref{eq2} and \eqref{eq3} we found that the normalization constant $A$ is given by
\begin{equation}
|A|^2=\frac{\eta q}{2\omega_k(2\pi)^2}.
\label{eq2.1}
\end{equation}
The complete normalized solution is, hence, written as
\begin{equation}
\Phi_k(x)=\left[\frac{\eta q}{2\omega_k(2\pi)^2}\right]^{\frac{1}{2}}e^{-i\omega_kt+i(n-\beta)q\varphi+i\nu z}J_{q|(n-\beta)|}(\eta r).
\label{eq4}
\end{equation}

The spectrum of quantum vacuum fluctuations of the scalar field in the quasi-periodically identified spacetime of a conical defect as considered here, when compared with the Minkowski spacetime, is definitely changed. The changes in the quantum vacuum fluctuations is, then, studied by means of the positive frequency Wightman  function defined  in the vacuum state $|0\rangle$ and given in terms of the product of the field operator $\hat{\Phi}(x)$ at different points, that is, $W(x,x')=\langle 0|\hat{\Phi}(x)\hat{\Phi}(x')|0\rangle$. The field operator once expanded in terms of the complete set of normalized mode functions, Eq. \eqref{eq4}, allows us to write the Wightman function in the form
\begin{eqnarray}
W(x,x^{\prime})&=&\sum_k\Phi_k(x)\Phi_k^*(x^{\prime})\nonumber\\
&=&\frac{q}{2(2\pi)^2}\sum_k\frac{e^{-i\omega_k\Delta t}}{\omega_k}e^{i\nu\Delta z}e^{i(n-\beta)q\Delta\varphi}\eta J_{q|(n-\beta)|}(\eta r)J_{q|(n-\beta)|}(\eta r^{\prime}),
\label{eq5}
\end{eqnarray}
where $\Delta t = t-t'$, $\Delta\varphi = \varphi - \varphi'$, $\Delta z = z-z'$ and we have used the notation
\begin{equation}
\sum_k=\int_{-\infty}^{\infty}d\nu\int_{0}^{\infty}d\eta\sum^{\infty}_{n=-\infty}.
\label{eq6}
\end{equation}
The procedure to perform the integrals and sum above requires that we first make a Wick rotation $\Delta\tau=i\Delta t$ in \eqref{eq5} and, then, use the identities (19) and (21) of Ref. \cite{Braganca:2019mvj}. This allows us to evaluate the integrals in $\nu$ and $\eta$ to obtain
\begin{eqnarray}
W(x,x^{\prime})=\frac{q}{2(2\pi)^2}\int_{0}^{\infty}\frac{ds}{s^3}e^{-s^2m^2-\frac{\Delta\zeta^2}{4s^2}}e^{-iq\beta\Delta\varphi}\sum_{n=-\infty}^{\infty}e^{inq\Delta\varphi}I_{q|(n-\beta)|}(rr'/2s^2),
\label{W1}
\end{eqnarray}
where $\Delta\zeta^2 = \Delta\tau^2 + \Delta z^2 + r'^2 + r^2$. Furthermore, the sum in $n$ above can be performed by using the summation formula (A.7) of Ref. \cite{deMello:2014ksa}. This makes possible to perform the integral in $s$, providing the closed form for the Wightman function\footnote{See Refs. \cite{Braganca:2019mvj, Braganca:2014qma, Mota:2017slg} for a similar procedure.}
\begin{eqnarray}
W(x,x^{\prime})&=&\frac{m^2}{(2\pi)^2}e^{-iq\beta \Delta\varphi}\left\{\sum_ne^{i\beta(2\pi n-q\Delta\varphi)}f_1(m\sigma_n)\right.\nonumber\\
\ &\ &\left.-\frac{q}{2i\pi}\sum_{j=+,-}je^{jiq\beta\pi}\int_{0}^{\infty}dy\frac{\cosh[qy(1-\beta)]-\cosh(q\beta y)e^{-iq(\Delta\varphi+j\pi)}}{\cosh(qy)-\cos[q(\Delta\varphi+j\pi)]}f_1(m\sigma_y)\right\},
\label{eq8}
\end{eqnarray}
where the function $f_{\mu}(x)$ is defined in terms of the Macdonald function $K_{\mu}(x)$ as
\begin{equation}
f_{\mu}(x)=\frac{K_{\mu}(x)}{x^{\mu}},
\label{eq8.2}
\end{equation}
and
\begin{subequations}
\begin{align}
\frac{\sigma_n^2}{2rr^{\prime}}&=\frac{\Delta\zeta^2}{2rr^{\prime}}-\cos\left(\frac{2n\pi}{q}-\Delta\varphi\right),
\label{eq8.3}\\
\frac{\sigma_y^2}{2rr^{\prime}}&=\frac{\Delta\zeta^2}{2rr^{\prime}}+\cosh y.
\label{eq8.4}
\end{align}
\end{subequations}
Note that the sum in $n$ in the first term on the r.h.s of \eqref{eq8} needs to be taken under the restriction \cite{Braganca:2019mvj, Braganca:2014qma, Mota:2017slg}
\begin{equation}
-\frac{q}{2}+\frac{\Delta\varphi}{\varphi_0}\leq n\leq\frac{q}{2}+\frac{\Delta\varphi}{\varphi_0}.
\label{eq8.1}
\end{equation}
Note also that in the case $q< 2$, the first term on the r.h.s of \eqref{eq8} only exists for $n=0$  \cite{Braganca:2019mvj, Braganca:2014qma, Mota:2017slg} . This term, in the coincidence limit $x'\rightarrow x$, diverges in the calculation of the physical observables and as such it should be subtracted to obtain the finite renormalized quantity.

It is evident that in the case the conical spacetime is not quasi-periodically identified ($\beta=0$) the expression \eqref{eq8} reduces to the known expression for the Wightman function in the pure conical spacetime \cite{BezerradeMello:2011nv, BezerradeMello:2011sm}, i.e,
\begin{eqnarray}
W(x,x^{\prime})&=&\frac{m^2}{(2\pi)^2}\left\{\sum_n
f_1(m\sigma_n)\right.\nonumber\\
\ &\ &\left.-\frac{q}{2\pi}\sum_{j=+,-}\int_{0}^{\infty}dy\frac{\sin(q\pi+jq\Delta\varphi)}{\cosh(qy)-\cos[q(\Delta\varphi+j\pi)]}f_1(m\sigma_y)\right\}.
\label{eq9}
\end{eqnarray}
Conversely, in the absence of a conical ($q=1$), in other words, in a quasi-periodically identified Minkowski spacetime, the Wightman function becomes
\begin{eqnarray}
W(x,x^{\prime})&=&\frac{m^2e^{-2i\beta\Delta\varphi}}{(2\pi)^2}\Biggl\{f_1(m\sigma_0)\nonumber\\
\ &\ &\left.-\frac{e^{i\beta\Delta\varphi}\sin(\pi\beta)}{\pi}\int_{0}^{\infty}dy\frac{\cosh[y(1-\beta)]+\cosh(\beta y)e^{-i\Delta\varphi}}{\cosh y+\cos\Delta\varphi}f_1(m\sigma_y)\right\}.
\label{eq10}
\end{eqnarray}
The first term on the r.h.s of the above expression is the term mentioned before that diverges in the coincidence limit $x'\rightarrow x$ $(\sigma_0\rightarrow 0)$. It is defined as the Hadamard function
\begin{eqnarray}
G_H(x,x^{\prime})=\frac{m^2e^{-2i\beta\Delta\varphi}}{(2\pi)^2}f_1(m\sigma_0),
\label{HF}
\end{eqnarray}
which should be subtracted in the calculation of the physical observables \cite{Christensen:1976vb, Wald:1978pj}. It is in fact the (divergent) Minkowski contribution to the Wightman function.

In the massless scalar field case we can obtain the Wightman function from Eq. \eqref{eq8} by taking the limit $m\rightarrow 0$. This gives
\begin{eqnarray}
W(x,x^{\prime})&=&\frac{e^{-i\beta q\Delta\varphi}}{(2\pi)^2}\left\{\sum_n\frac{e^{i\beta(2\pi n-q\Delta\varphi)}}{\sigma_n^2}-\frac{q}{2i\pi}\sum_{j=+,-}je^{ij\pi q\beta}\right.\nonumber\\
\ &\ &\times\left.\int_{0}^{\infty}dy\frac{\cosh[qy(1-\beta)]-\cosh(q\beta y)e^{-iq(\Delta\varphi+j\pi)}}{[\cosh(qy)-\cos[q(\Delta\varphi+j\pi)]]\sigma^2_y}\right\}.
\label{eq11}
\end{eqnarray}
It is straightforward to show that the massless case of Eqs. \eqref{eq9}, \eqref{eq10} and \eqref{HF} follows from Eq. \eqref{eq11}, showing the consistency of our results.

The exact closed expression for the Wightman function, Eq. \eqref{eq8}, in a quasi-periodically identified conical spacetime will allow us, in the next section, to study the physical observables we are interested in, namely, the Vacuum Expectation Value (VEV) of the field squared, the VEV of the energy-momentum tensor and the induced current density.

\section{Physical observables}
\label{secIII}
\subsection{Vacuum expectation value of $\Phi^2(x)$}
Formally, the standard prescription to calculate the VEV of the field squared\footnote{From now on we will use the short notation $\langle 0|\hat{O}|0\rangle = \langle\hat{O}\rangle$.}, $\langle\Phi^2\rangle$, is to take the Wightman function \eqref{eq8} in the coincidence limit $x'\rightarrow x$  \cite{Braganca:2019mvj, Braganca:2014qma, Mota:2017slg} , that is,
\begin{eqnarray}
\langle\Phi^2\rangle=\lim_{x^{\prime}\to x}W(x,x^{\prime}).
\label{eq27}
\end{eqnarray}
However, in order to perform this procedure we need first to subtract the divergent part described by the Hadamard function \eqref{HF}. In the coincidence limit we, thus, have the renormalized VEV of the field squared
\begin{eqnarray}
\langle\Phi^2\rangle_{\text{ren}}&=&\lim_{x^{\prime}\to x}[W(x,x^{\prime})-G_H(x,x^{\prime})]\nonumber\\
&=&\frac{2m^2}{(2\pi)^2}\left\{\sum_{n=1}^{[q/2]}\!^{*}\cos(2\pi n\beta)f_1(2mrs_n)-\frac{q}{2\pi}\int_{0}^{\infty}dyM(y,\beta,q)f_1(2mrs_y)\right\},
\label{eq27}
\end{eqnarray}
where $s_n=\sin(n\pi/q)$, and $s_y=\cosh(y/2)$ and
\begin{equation}
M(y,\beta,q) = \frac{\sin[q\pi(1-\beta)]\cosh(q\beta y) + \cosh[qy(1-\beta)]\sin(q\beta\pi)}{\cosh(qy)-\cos(q\pi)}.
\label{eq29}
\end{equation}
Note that $[q/2]$ represents the integer part of $q/2$ and the symbol $(*)$ present in the sign of summation in $n$ means that in the case $q$ is an integer number the sum in $n$ should be replaced with  \cite{Braganca:2019mvj, Braganca:2014qma, Mota:2017slg, BezerradeMello:2011nv, BezerradeMello:2011sm}
\begin{eqnarray}
\sum_{n=1}^{[q/2]}\!^{*}\rightarrow \frac{1}{2}\sum_{n=1}^{q-1}.
\label{sr}
\end{eqnarray}

The massless scalar field case is obtained from \eqref{eq27} by taking the limit $m\rightarrow 0$. This provides
\begin{eqnarray}
\langle\Phi^2\rangle_{\text{ren}}&=&\frac{1}{8\pi^2r^2}\left\{\sum_{n=1}^{[q/2]}\!^{*}\frac{\cos(2\pi\beta n)}{\sin^2(n\pi/q)}-\frac{q}{2\pi}\int_{0}^{\infty}dy\frac{M(y,\beta,q)}{\cosh^2(y/2)}\right\}.
\label{eq33}
\end{eqnarray}
Hence, the fluctuations of the quantum vacuum of the scalar field do not average to zero as we can clearly see through the expressions for the renormalized VEV of the field squared in the massive and massless scalar field cases given by Eqs. \eqref{eq27} and \eqref{eq33}, respectively. It is clear in both of these expressions that taking $q=1$ and $\beta =0$ the renormalized VEV of the field squared average to zero, as it should be. It is also clear that, in the case $\beta=0$, the results in \eqref{eq27} and \eqref{eq33} recover the VEV of the field squared in a conical spacetime (cosmic string or dispiration)  \cite{BezerradeMello:2011nv, BezerradeMello:2011sm}. Conversely, in the case $q=1$, the first term on the r.h.s of Eqs. \eqref{eq27} and \eqref{eq33} are absent and we obtain the VEV of the field squared in a quasi-periodically identified Minkowski spacetime. It is worth pointing out that, in the limit $mr\gg 1,$ the VEV of the field squared in the massive case \eqref{eq27} is exponentially suppressed. In the opposite limit, $mr\ll 1$, Eq. \eqref{eq27}  provides the VEV of the field squared in the massless case, Eq. \eqref{eq33}.

It is interesting to analyse the case $\beta =1/2$ (twisted scalar field) separately for the renormalized VEV of the field squared in Eq. \eqref{eq33}. The exact expression for any value of $q$ is given by
\begin{eqnarray}
\langle\Phi^2\rangle_{\text{ren}}&=&\frac{1}{8\pi^2r^2}\left\{\sum_{n=1}^{[q/2]}\!^{*}\frac{(-1)^n}{\sin^2(n\pi/q)}-\frac{q}{2\pi}\int_{0}^{\infty}dy\frac{M(y,1/2,q)}{\cosh^2(y/2)}\right\}.
\label{eq33tw}
\end{eqnarray}
One should remind, however, that for $q<2$ the first term on the r.h.s. is absent. 

In particular, if the conical parameter $q$ is an integer number, the expression \eqref{eq33tw} can be simplified by noting Eq. \eqref{sr}. This provides an algebraic expression for the sum in $n$ on the r.h.s of \eqref{eq33tw}, that is,
\begin{eqnarray}
\frac{1}{2}\sum_{n=1}^{q-1}\frac{(-1)^n}{\sin^2(n\pi/q)}=-\frac{1}{12}(q^2 + 2),\qquad\text{for}\qquad q-\text{even}.
\label{twisted1}
\end{eqnarray}
Note that the sum in $n$ above is zero for $q$-odd. Hence, for even values of $q$, in the twisted massless scalar field case, from Eq. \eqref{eq33tw}, we have
\begin{eqnarray}
\langle\Phi^2\rangle_{\text{ren}}&=&-\frac{1}{96\pi^2r^2}(q^2+2),
\label{twistedeven}
\end{eqnarray}
where $M(y,1/2,q)=0$ for even $q$. The expression in Eq. \eqref{twistedeven} was obtained previously in Ref. \cite{smith} for any value of $q$. In fact, an analytic continuation can be assumed for the result in Eq. \eqref{twistedeven} to be valid for any value of $q$. A numerical analysis of \eqref{twistedeven} comparing it with the actual general expression \eqref{eq33tw}, valid for any value of $q$, shows that the analytical continuation of \eqref{twistedeven} to be valid for any value of $q$ is plausible and agrees with \eqref{eq33tw} whatever the value of $q$ we take. 

On the other hand, by considering only odd values of $q$, the only contribution to the VEV of the field square comes from the second term on the r.h.s of \eqref{eq33tw}. The expression is then written as 
\begin{eqnarray}
\langle\Phi^2\rangle_{\text{ren}}&=&-\frac{q\sin(q\pi/2)}{16\pi^3r^2}\int_{0}^{\infty}dy\frac{[\cosh(qy/2)]^{-1}}{\cosh^2(y/2)}, \qquad\text{for}\qquad q-\text{odd}.
\label{twisted2}
\end{eqnarray}
Note that an analytical continuation can not be assumed for the expression above to be valid for any value of $q$ since it does not reproduce the values of \eqref{twistedeven} for any value of $q$.

In the case there is no conical defect in Eq. \eqref{twisted2}, that is, $q=1$, we obtain that the resulting integral in $y$ is given by $\frac{\pi}{2}$, leading to
\begin{eqnarray}
\langle\Phi^2\rangle_{\text{ren}}&=&-\frac{1}{32\pi^2r^2}.
\label{twisted3}
\end{eqnarray}
which agrees with Refs. \cite{PhysRevD.21.949, smith}. Note that we obtain the same result \eqref{twisted3}  by making $q=1$ in \eqref{twistedeven}, as it should be. Therefore, our $\beta$-dependent general expressions \eqref{eq27} and \eqref{eq33} for the VEV of the field squared for the massive and massless scalar fields recover the long known results \eqref{twistedeven} and \eqref{twisted3} \cite{PhysRevD.21.949, smith}.
%
\subsection{VEV of the energy-momentum tensor}
%
Let us now turn to the calculation of the VEV of the energy-momentum tensor arising as a consequence of the quasi-periodically identified conical spacetime. We can do that by considering the expression of the VEV of the energy-momentum tensor for a charged scalar field derived in \cite{Braganca:2019mvj, Santos:2019ais}. This expression, when assuming the absence of a four-vector gauge field, is given by
\begin{eqnarray}
\langle T_{\mu\nu}\rangle=2\lim_{x^{\prime}\to x}\partial_{\mu^{\prime}}\partial_{\nu}W(x,x^{\prime})+2\left[\left(\xi-\frac{1}{4}\right)g_{\mu\nu}\Box-\xi\nabla_{\mu}\nabla_{\nu}-\xi R_{\mu\nu}\right]\langle\Phi^2\rangle,
\label{eq35}
\end{eqnarray}
where $R_{\mu\nu}$ is the Ricci tensor and $\Box$ is the box or d'Alembertian operator. Moreover, by using Eq.  \eqref{eq27} we are able to obtain $\Box\langle\Phi^2\rangle$, that is,
\begin{eqnarray}
\Box\langle\Phi^2\rangle_{\text{ren}}&=&-\frac{8m^4}{(2\pi)^2}\left\{\sum_{n=1}^{[q/2]}\!^{*}\cos(2\pi n\beta)\left[(2mr)^2s_n^4f_3(2mrs_n)-2s_n^2f_2(2mrs_n)\right]\right.\nonumber\\
\ &\ &-\frac{q}{2\pi}\int_0^{\infty}dyM(y,\beta,q)\left[(2mr)^2s_y^4f_3(2mrs_y)-2s_y^2f_2(2mrs_y)\right]\Biggl\}.
\label{eq38}
\end{eqnarray}
%
\begin{figure}[!htb]
\begin{center}
\includegraphics[width=0.4\textwidth]{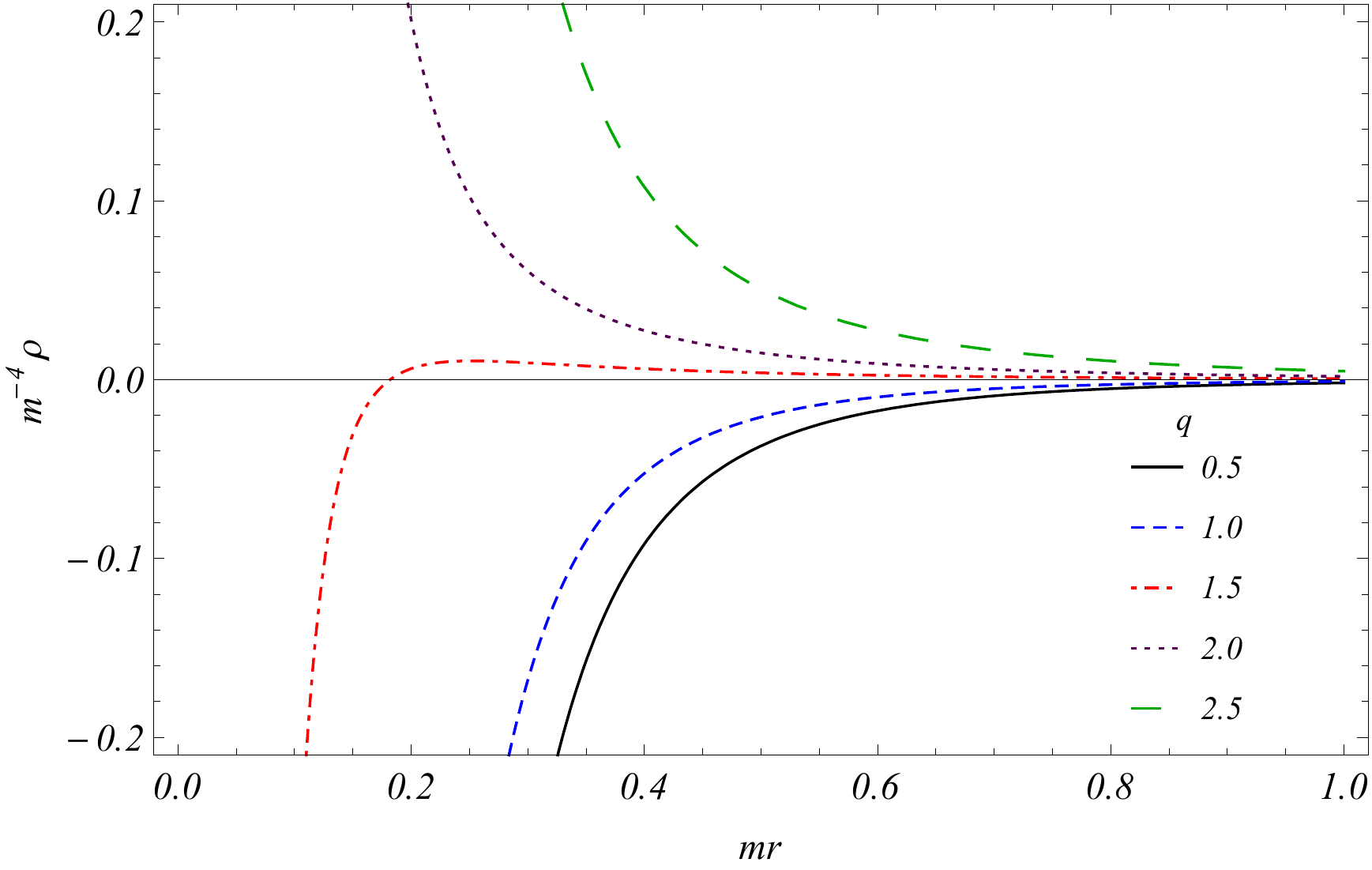}
%
\includegraphics[width=0.405\textwidth]{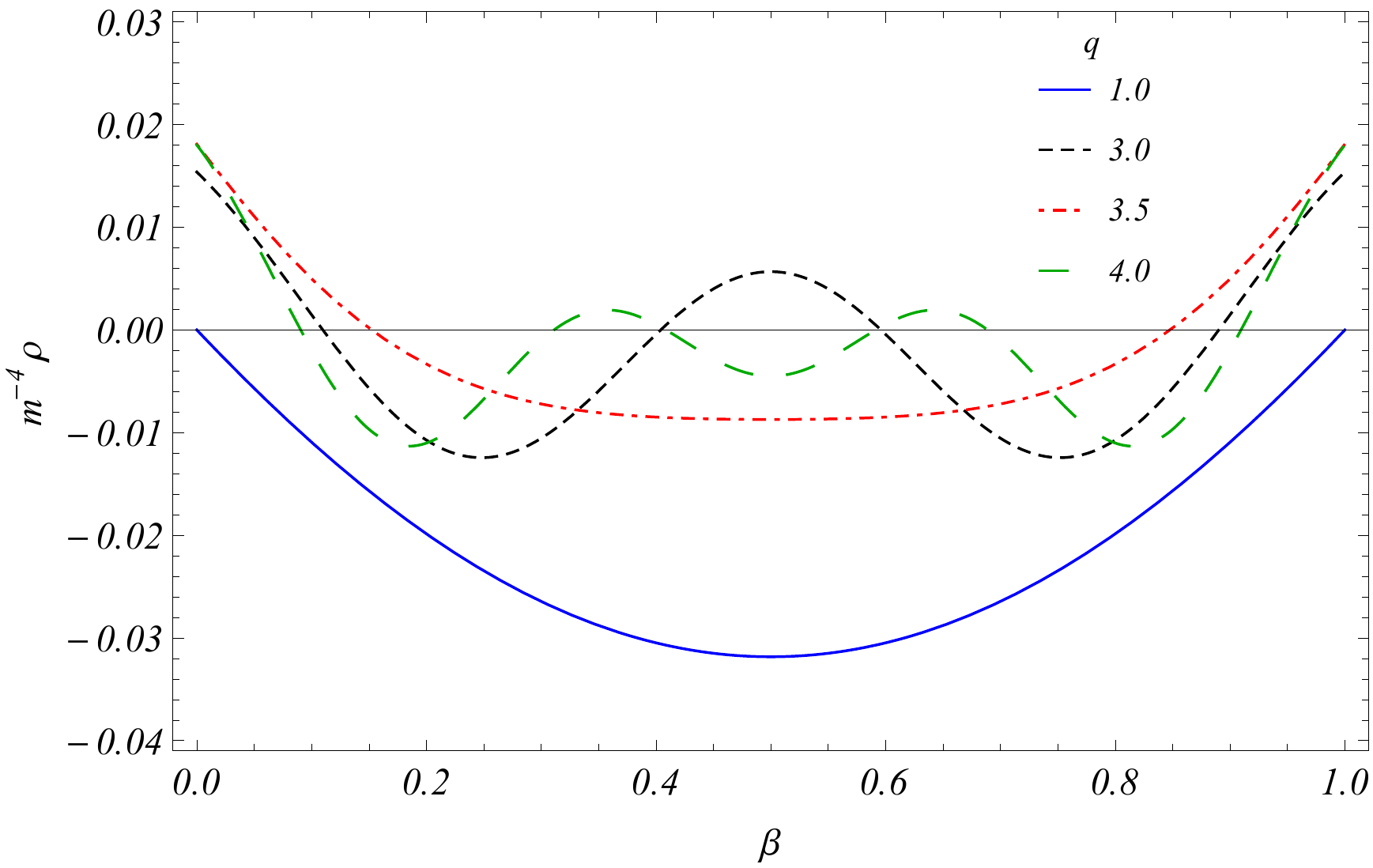}
\caption{Renormalized massive scalar vacuum energy density, $\rho=\langle T^{0}_{0}\rangle_{\text{ren}}$, in units of $m^4$, plotted in terms of $mr$ (on the left) and $\beta$ (on the right). Note that  $\beta=1/11$ value has been taken for the plot on the left and $mr = 1$ for the plot on the right.\small}
\label{figure1}
\end{center}
\end{figure}
%
Note that the only derivative contribution in the d'Alembertian operator is the one due to the radial component, $r$. Hence, from Eqs. \eqref{eq35}, \eqref{eq38} and \eqref{eq8} for the minimally coupled case, $\xi=0$, we find that the renormalized VEV of the energy-momentum tensor for a massive charged scalar field is found to be
\begin{eqnarray}
\langle T^{\mu}_{\nu}\rangle_{\text{ren}}=\frac{m^4}{\pi^2}\left\{\sum_{n=1}^{[q/2]}\!^{*}\cos(2\pi n\beta)F^{\mu}_{\nu}(2mr,s_n)-\frac{q}{2\pi}\int_0^{\infty}dyM(y,\beta,q)F^{\mu}_{\nu}(2mr,s_y)\right\},
\label{eq40}
\end{eqnarray}
where the functions $F^{\mu}_{\nu}(u,s)$ are defined as
\begin{eqnarray}
F^0_{0}(u,s) &=&u^2s^4f_3(us)-[2s^2+1]f_2(us),\nonumber\\
F^1_{1}(u,s) &=&-f_2(us),\nonumber\\
F^2_{2}(u,s) &=&u^2s^2f_3(us) - f_2(us),\nonumber\\
F^3_{3}(u,s) &=&F^0_{0}(u,s).
\label{funF}
\end{eqnarray}
Some observations are worth to be pointed out at this stage. Firstly, we can notice that the VEV of the energy-momentum tensor \eqref{eq40} presents a boost invariance in the $z$ direction which leads to $F^3_{3}(u,s) =F^0_{0}(u,s)$. Moreover, in order to calculate the (2,2)-component of the VEV of the energy-momentum tensor, \eqref{eq40}, we made use of Eq. (52) from \cite{BezerradeMello:2011sm}. Note also that, in the absence of a conical defect $(q=1)$, there is still a nonzero contribution to the VEV of the energy-momentum tensor coming from the second term on the r.h.s of Eq. \eqref{eq40} due to the quasi-periodic condition \eqref{eq1}. On the other hand, in the absence of the quasi-periodic condition \eqref{eq1}, we recover the VEV of the energy-momentum tensor generated by the presence of a conic defect\footnote{In fact, we do not recover exactly the same expression as in Refs. \cite{BezerradeMello:2011nv, BezerradeMello:2011sm}, that considered a real scalar field. It is the expression obtained in the latter multiplied by two that we recover. This factor of two comes from \eqref{eq35} which is the definition of the VEV of the energy-momentum tensor of a charged scalar field.} \cite{BezerradeMello:2011nv, BezerradeMello:2011sm}. The VEV of the energy-momentum tensor, in the limit $mr\gg 1$, is exponentially suppressed while in the opposite limit, $mr\ll 1$, provides the massless expression as the dominant contribution (see Eq. \eqref{massless}). Evidently, when $r\rightarrow 0$, keeping $m\neq 0$, the energy-momentum tensor diverges. These asymptotic behaviours are shown in the left plot in Fig.\ref{figure1} for the (0,0)-component (energy density), considering different values of the conical parameter $q$. It is clear that for a given value of $\beta$, the sign of the energy density changes as the value of $q$ increases. The plot on the right of Fig.\ref{figure1} for the energy density is in terms of $\beta$ and shows that it begins to oscillates as the value of $q$ increases. Finally, we have proved that the VEV of the energy-momentum tensor satisfies both the covariant conservation condition $\nabla_{\mu}\langle T^{\mu}_{\nu}\rangle_{\text{ren}}=0$ and the trace identity for a massive charged scalar field, i.e.,
\begin{eqnarray}
\langle T^{\mu}_{\mu}\rangle_{\text{ren}}=6(\xi-1/6)\nabla_{\mu}\nabla^{\mu}\langle\Phi^2\rangle_{\text{ren}} + m^2\langle\Phi^2\rangle_{\text{ren}}.
\label{trace}
\end{eqnarray}
%
\begin{figure}[!htb]
\begin{center}
\includegraphics[width=0.4\textwidth]{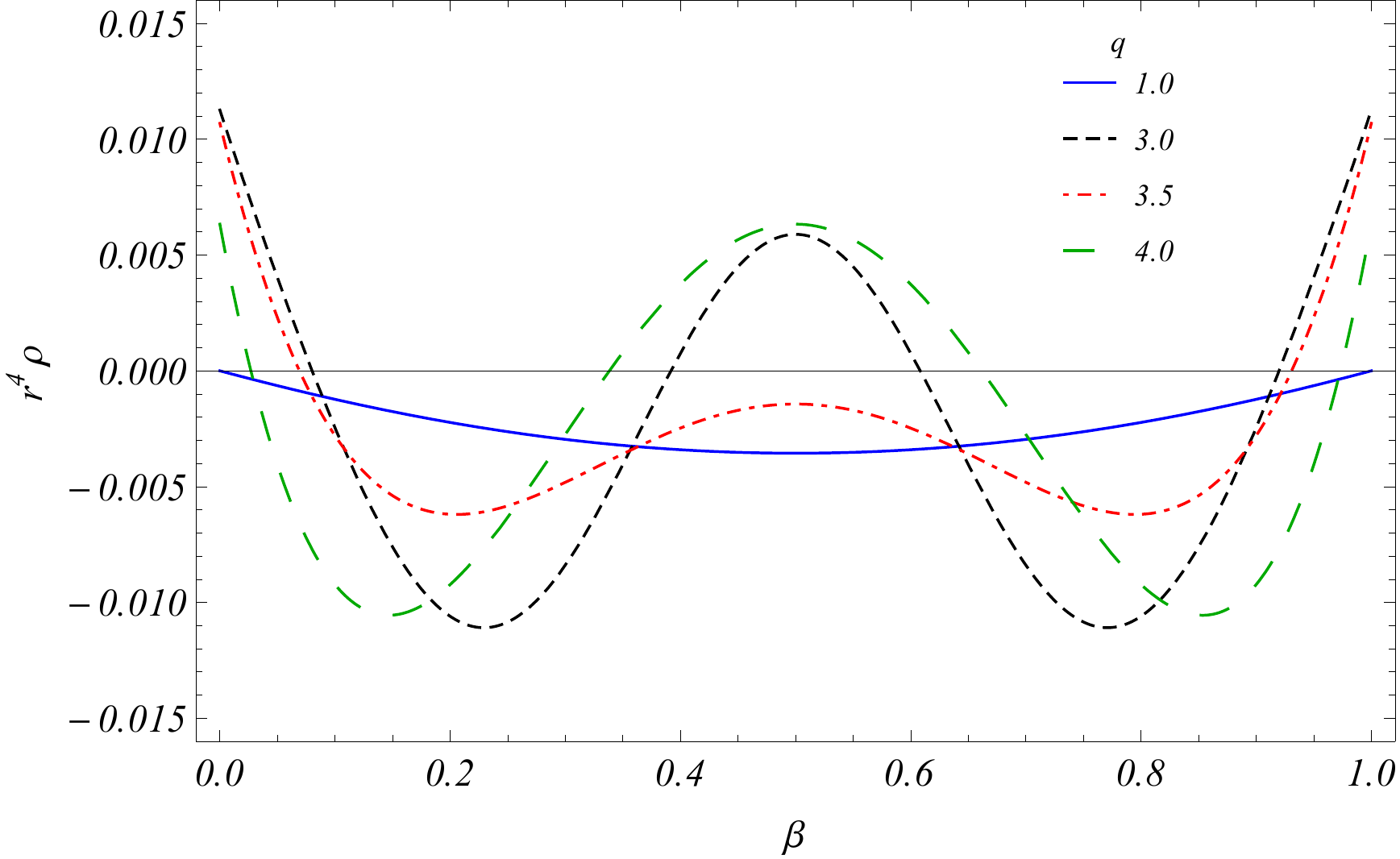}
\caption{Renormalized massless scalar vacuum energy density, $\rho=\langle T^{0}_{0}\rangle_{\text{ren}}$, in units of $r^{-4}$, plotted in terms of $\beta$.\small}
\label{figure2}
\end{center}
\end{figure}
%

The renormalized VEV of the energy-momentum tensor in the massless scalar field case is obtained exactly by taking the limit $m\rightarrow 0$ of the massive case, Eq. \eqref{eq40}. This provides
\begin{eqnarray}
\langle T^{\mu}_{\nu}\rangle_{\text{ren}}=\frac{1}{16\pi^2r^4}\left\{\sum_{n=1}^{[q/2]}\!^{*}\cos(2\pi n\beta)H^{\mu}_{\nu}(s_n)-\frac{q}{2\pi}\int_0^{\infty}dyM(y,\beta,q)H^{\mu}_{\nu}(s_y)\right\},
\label{massless}
\end{eqnarray}
where the functions $H^{\mu}_{\nu}(s)$ are defined as
\begin{eqnarray}
H^{0}_{0}(s) &=&\frac{4}{s^2}-\frac{2}{s^4},\nonumber\\
H^{1}_{1}(s) &=&-\frac{2}{s^4},\nonumber\\
H^{2}_{2}(s) &=&\frac{6}{s^4},\nonumber\\
H^{3}_{3}(s) &=&H^{0}_{0}(s).
\label{funF}
\end{eqnarray}
As we can see, the VEV of the energy-momentum tensor in the massless scalar field case, \eqref{massless}, is also boost invariant in the $z$-direction since $H^{3}_{3}(s) = H^{0}_{0}(s)$. It is worth noticing that the expression \eqref{massless} vanishes for very large radial distances and diverges as $r$ goes to zero, in both cases with a power of $r^{-4}$. In order to see the behaviour of \eqref{massless} with respect to $\beta$ its plot is shown in Fig.\ref{figure2} and it is very similar to the massive case, that is, as the conic parameter is increased, the VEV of the energy-momentum tensor in the massless scalar field case start to oscillate more. Moreover, when $\beta=0$, our result is consistent with the expression for the energy-momentum tensor purely as a consequence of a conic defect found in Refs. \cite{BezerradeMello:2011nv, BezerradeMello:2011sm}, for a real scalar field. Conversely, in the absence of a conic defect, that is, $q=1$, the only contribution comes from the second term on the r.h.s of \eqref{massless}.

As we have done in the previous section for the VEV of the field squared, let us also consider separately the twisted scalar field case, i.e., when $\beta=1/2$. Again, for simplicity, we will take integer values of $q$ and analyse the first term on the r.h.s of \eqref{massless}. In this case, by making use of \eqref{sr}, we get
\begin{eqnarray}
\frac{1}{2}\sum_{n=1}^{q-1}(-1)^nH^{\mu}_{\nu}(s_n) = \left[\frac{7q^4 + 8}{360}(1, 1,-3,1)-\frac{q^2+2}{9}(2,-1, 3, 2)\right],\qquad\text{for}\qquad q-\text{even},
\label{emtwisted1}
\end{eqnarray}
where the numbers in parentheses stand for the diagonal elements. Thus, the VEV of the energy-momentum tensor, in this case, is given by
\begin{eqnarray}
\langle T^{\mu}_{\nu}\rangle_{\text{ren}}=\frac{1}{16\pi^2r^4}\left[\frac{7q^4 + 8}{360}(1, 1,-3,1)-\frac{q^2+2}{9}(2,-1, 3, 2)\right],
\label{masslesstw}
\end{eqnarray}
where  $M(y,1/2,q)=0$ for even $q$. Although \eqref{masslesstw} has been obtained by using the result \eqref{emtwisted1}, valid for even $q$, an analytic continuation can be assumed in order for it to be valid for any value of $q$, in the same way we have done for the expression \eqref{twistedeven}. Indeed, taking any value of $q$, we can numerically verify that Eq. \eqref{masslesstw} is consistent with Eq. \eqref{massless}, for $\beta=1/2$. The result \eqref{masslesstw} is also consistent with the one obtained in Ref.  \cite{smith}, for a real scalar field.

In the absence of a conical defect ($q=1$), the contribution is solely due to the massless and charged twisted scalar field and can be obtained from \eqref{masslesstw} as
\begin{eqnarray}
\langle T^{\mu}_{\nu}\rangle_{\text{ren}} = \frac{1}{128\pi^2r^2}(-5, 3,-9,-5).
\label{emtwisted3}
\end{eqnarray}
The expression above is also consistent with the result obtained in Refs. \cite{PhysRevD.21.949, smith} for a real twisted scalar field in Minkowski spacetime.

\subsection{Induced current density}
A complex scalar field obeying a quasi-periodicity condition \eqref{eq1} in a conical spacetime has also an induced current density as a result of its quantum vacuum fluctuations be modified. Formally, the VEV of the induced current density is obtained as \cite{Braganca:2014qma}
\begin{equation}
\langle J_{\mu}\rangle=i\lim_{x^{\prime}\to x}(\partial_{\mu}-\partial_{\mu^{\prime}})W(x,x^{\prime}).
\label{eq12.0}
\end{equation}
This induced current density arises in fact due to the action \eqref{action} be invariant under the quasi-periodicity condition \eqref{eq1}. Since the latter affects only the $\varphi$-coordinate, the existing action symmetry induces only a nonzero azimuthal current density, according to the Noether's theorem.

In order to calculate the $\varphi$-component of the VEV of the current density \eqref{eq12.0}, the better approach is to take first the $\varphi$-derivative of \eqref{W1} and, then, take the coincidence limit $x'\rightarrow x$. In this case we have

\begin{eqnarray}
\langle J_{\varphi}\rangle=-\frac{q}{(2\pi)^2}\int_{0}^{\infty}\frac{ds}{s^3}e^{-s^2m^2-\frac{r^2}{2s^2}}\sum_{n=-\infty}^{\infty}q(n-\beta)I_{q|(n-\beta)|}(r^2/2s^2).
\label{ICD}
\end{eqnarray}
The sum in $n$ present in the above expression has been worked out in Ref. \cite{Braganca:2014qma} (Eq. A19). By making use of the result from the latter we are able to perform the integral in $s$ to obtain
%
%
\begin{figure}[!htb]
\begin{center}
\includegraphics[width=0.41\textwidth]{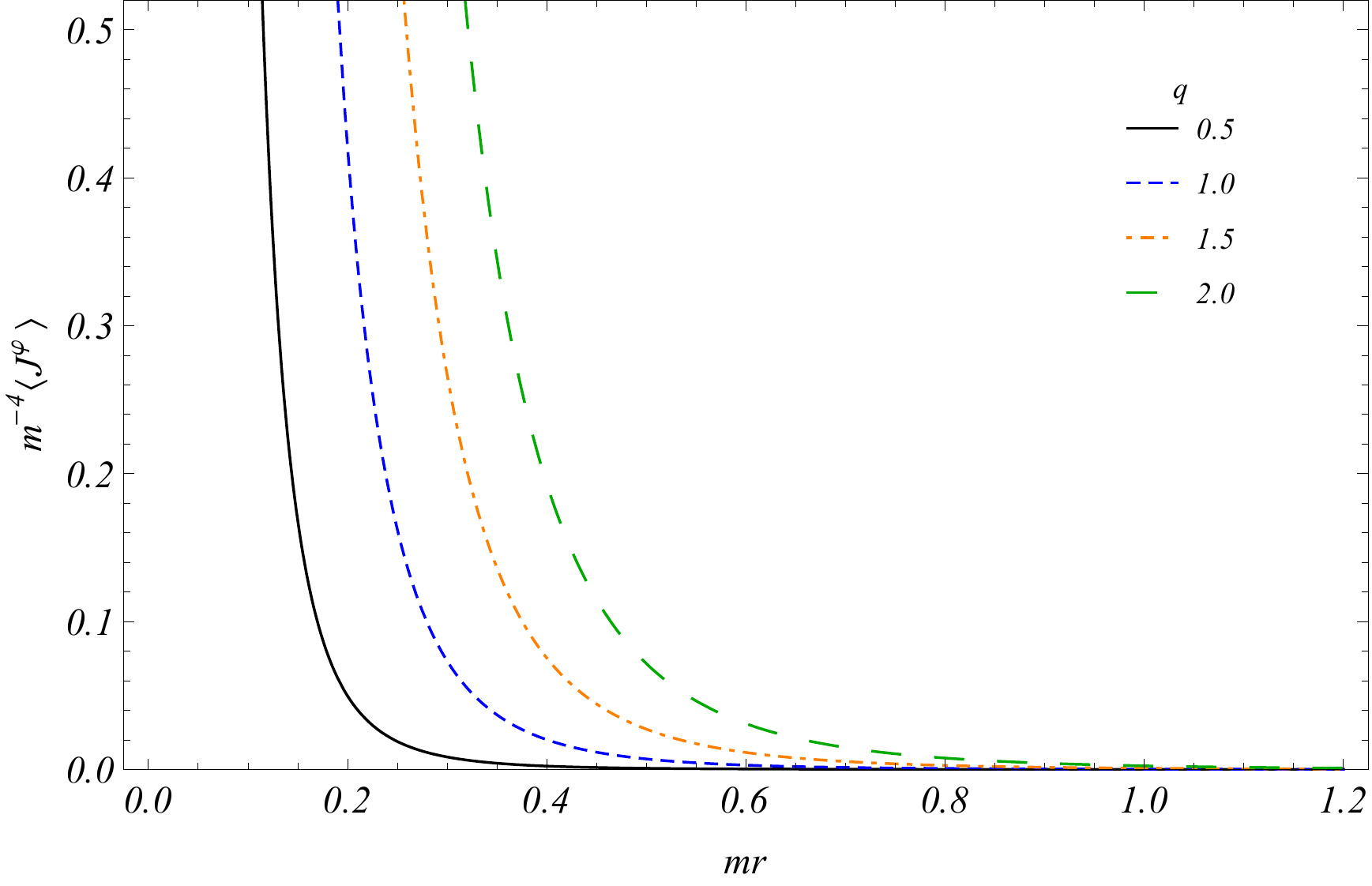}
%
\includegraphics[width=0.4333\textwidth]{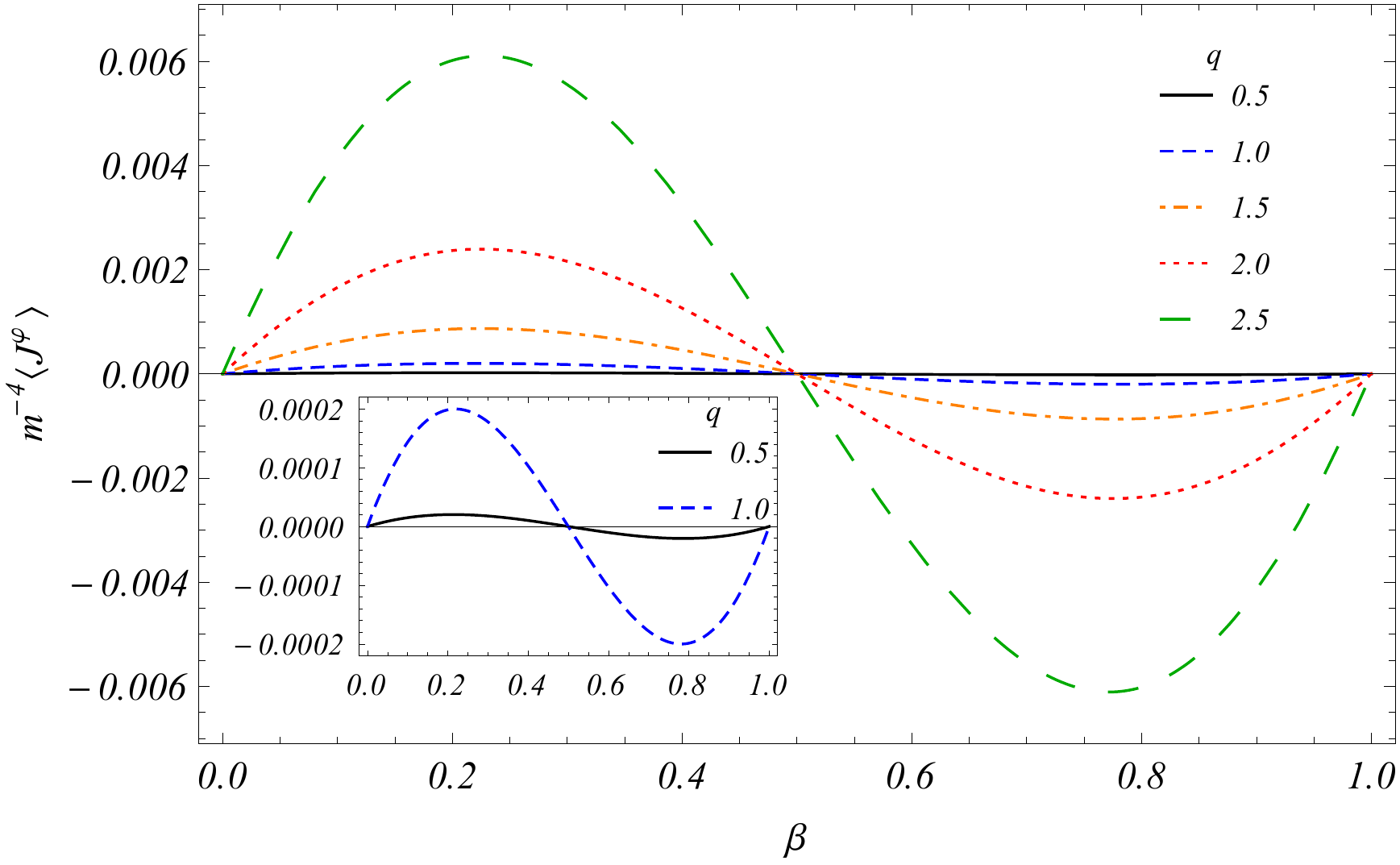}
\caption{Renormalized induced current density, \eqref{eq19}, in units of $m^{4}$, plotted in terms of $mr$ (on the left) and $\beta$ (on the right). Note that $\beta=1/6$ has been taken for the plot on the left and $mr = 1$ for the plot on the right.\small}
\label{figure3}
\end{center}
\end{figure}
%
%
\begin{eqnarray}
\langle J^{\varphi}\rangle&=&\frac{m^4}{\pi^2}\left\{\sum_{k=1}^{[q/2]}\!^{*}\sin\left(\frac{2k\pi}{q}\right)\sin(2k\pi\beta)f_2(2mrs_k)+\frac{q}{2\pi}\int_{0}^{\infty}dyg(y,\beta,q)f_2(2mrs_y)\right\},
\label{eq19}
\end{eqnarray}
where
\begin{equation}
g(y,\beta,q)=\frac{\sin(q\pi\beta)\sinh[(1-|\beta|)qy]-\sinh(yq\beta)\sin[(1-|\beta|)q\pi]}{[\sinh(y)]^{-1}\left[\cosh(qy)-\cos(q\pi)\right]}.
\label{eq16}
\end{equation}
The induced current density \eqref{eq19} associated with a massive and charged scalar field is exponentially suppressed in the regime $mr\gg1$ while it provides as the main leading contribution the expression for the massless and charged scalar field in the regime $mr\ll 1$. Furthermore, if we keep $m$ fixed and take $r\rightarrow 0$, the induced current density  \eqref{eq19} diverges. These asymptotic behaviours are shown in the left plot of Fig.\ref{figure3}. It also shows that the induced current density is enhanced as $q$ is increased. On the right plot of Fig.\ref{figure3}, on the other hand, for a fixed value of $mr$, the dependence of the induced current density on $\beta$ makes it oscillate more as $q$ is increased. The plot on the right of Fig.\ref{figure3} also shows that the induced current density is positive when $\beta<0.5$ and negative for $\beta>0.5$, vanishing when $\beta=0.5$. That is, there is no induced current density for a twisted scalar field.

\begin{figure}[!htb]
\begin{center}
\includegraphics[width=0.4\textwidth]{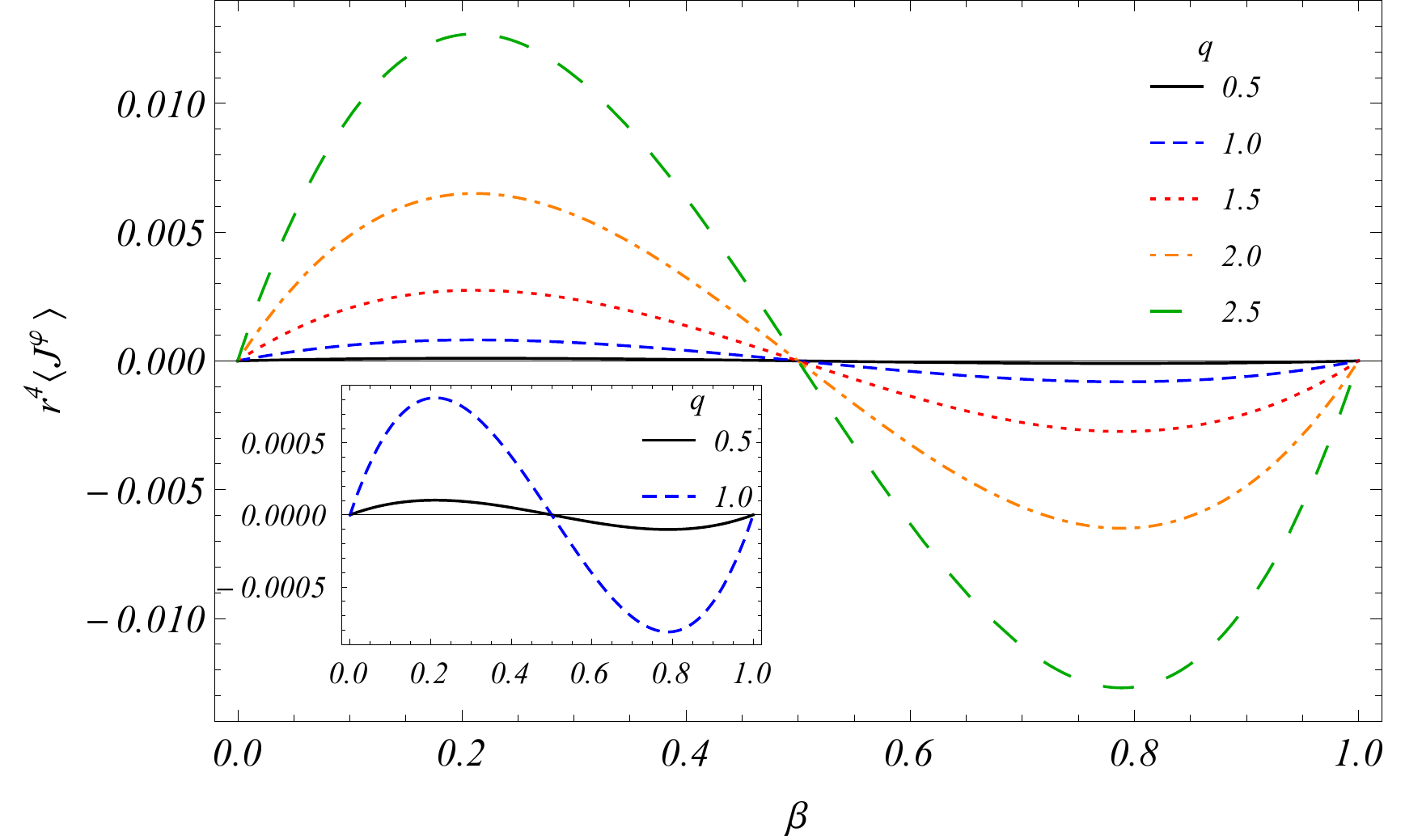}
\caption{Renormalized induced current density, \eqref{eq24}, in units of $r^{-4}$, plotted in terms of $\beta$.\small}
\label{figure4}
\end{center}
\end{figure}

Now, if one takes the limit $m\rightarrow0$ in \eqref{eq19}, we find the exact closed expression for the induced current density of a massless and charged scalar field
\begin{eqnarray}
\langle J^{\varphi}\rangle&=&\frac{1}{8\pi^2 r^4}\left\{\sum_{k=1}^{[q/2]}\!^{*}\sin\left(\frac{2k\pi}{q}\right)\frac{\sin(2k\pi\beta)}{\sin^4(k\pi/q)}+\frac{q}{2\pi}\int_{0}^{\infty}dy\frac{g(y,\beta,q)}{\cosh^4(y/2)}\right\}.
\label{eq24}
\end{eqnarray}
It is straightforward to see that this expression goes to zero as $r\rightarrow\infty$ and diverges when $r\rightarrow 0$, in both cases with power of $r^{-4}$. In Fig.\ref{figure4} the induced current density is plotted in terms of $\beta$ and shows that the more $q$ is increased the more the induced current density oscillates. It also shows, again, that the induced current density is positive when $\beta<0.5$ and negative for $\beta>0.5$, vanishing in the twisted scalar case $\beta=0.5$.
%
%

One should note that, from Eqs. \eqref{eq19} and \eqref{eq24}, if we take $\beta=0$ there is no current density induced solely by the conical spacetime, neither there is an induced current density in the case of a twisted scalar field, $\beta=1/2$. On the other hand, if we consider a quasi-periodically identified Minkowski spacetime, that is, $q=1$, we have
\begin{equation}
\langle J^{\varphi}\rangle=\frac{m^4}{2\pi^3}\int_{0}^{\infty}dyg(y,\beta,1)f_2(2mrs_y),
\label{eq23}
\end{equation}
which is the pure contribution induced by the quasi-periodicity condition \eqref{eq1}. The massless scalar field case follows from \eqref{eq23}, in the limit $m\rightarrow 0$, and is given by
\begin{eqnarray}
\langle J^{\varphi}\rangle=\frac{\sin(\pi\beta)}{32\pi^3r^4}\int_{0}^{\infty}dy\frac{\sinh(y)[\sinh[(1-\beta)y]-\sinh(y\beta)]}{\cosh^6(y/2)}.
\label{eq26}
\end{eqnarray}
An approximated expression can be obtained for Eq. \eqref{eq26} by noting that the main contribution for the integral to converge comes from $\cosh^6(y/2)$ present in the denominator. Thus, the term in the numerator can be expanded up to second order to provide
\begin{eqnarray}
\langle J^{\varphi}\rangle&\simeq&\frac{(1-2\beta)\sin(\pi\beta)}{32\pi^3r^4}\int_{0}^{\infty}dy\frac{y^2}{\cosh^6(y/2)}\nonumber\\
&\simeq&\frac{(1-2\beta)\sin(\pi\beta)}{180\pi^3r^4}\left(2\pi^2-15\right),
\label{eq26.2}
\end{eqnarray}
which is consistent with the induced current density being averaged to zero when $\beta=0$ and $\beta=1/2$, as it can be seen in Eqs. \eqref{eq19} and \eqref{eq24}-\eqref{eq26}.

It is not unexpected that a current density is induced since we have imposed that the charged scalar field has to obey the condition \eqref{eq1}. The action \eqref{action} is symmetric under \eqref{eq1} and this makes possible to obtain the results \eqref{eq19}, \eqref{eq24}, \eqref{eq23} and \eqref{eq26}. These are new results in light of the particular physical system we have considered, that is, a massive and charged scalar field propagating in a conical spacetime under restriction \eqref{eq1}. The latter is in fact a quasi-periodic condition which, as explained previously in section \ref{intro}, can mimic the properties of nanotubes. In this particularly interesting scenario, with potential applications, we also obtained the expressions for the VEV's of the field squared \eqref{eq27} and energy-momentum tensor \eqref{eq19} in the massive and charged scalar field case, which are more general than the results from Refs. \cite{PhysRevD.21.949, smith}. 
%
\section{Conclusion}
\label{secIV}

We have considered a quasi-periodic boundary condition, having as motivation its application in carbon nanotubes, being satisfied by a massive charged scalar field whose modes propagates in (3+1)-dimensions conical spacetime which could be either a disclination, where the conical parameter $q$ can assume any value above zero, or a cosmic string with $q\geq 1$. Under these conditions, the Klein-Gordon equation is solved for the scalar field and the complete normalized solution \eqref{eq4} is shown to depend on, $q$, and, $\beta$, the parameter characterizing the quasi-periodic condition \eqref{eq1}. The complete normalized solution is then used to obtain an exact and analytical expression for the Wightman function, in both the massive and massless charged scalar field cases, that is, Eqs. \eqref{eq8} and \eqref{eq11}.

The closed expressions for the Wightman function found provided the means to obtain how physical observables such as the field squared, energy-momentum tensor and current density are averaged to nonzero values. This has been clearly shown in the expressions \eqref{eq27} and \eqref{eq33} for the VEV of the field squared, in the expressions \eqref{eq40} and \eqref{massless} for the VEV of the energy-momentum tensor and in the expressions \eqref{eq19} and \eqref{eq24} for the induced current density. All these expressions have been obtained in the massive and massless scalar field cases and shown to depend on both the conical and quasi-periodic parameters $q$ and $\beta$, respectively. The expressions for the energy density, that is, the 00-component of the energy-momentum tensor and for the induced current density have also been plotted, in terms of $mr$ and $\beta$, in Figs.\ref{figure1}-\ref{figure4}.

The modification on the quantum vacuum fluctuations of the charged scalar field caused by the quasi-periodic condition and conical parameter certainly averages the physical observables analyzed to nonzero values. We have checked that in the case there is no quasi-periodic condition, that is, $\beta=0$, our results are in agreement with previous results found in literature. We have noted, however, that in this case the induced current density averages to zero which tell us that the conical defect by itself does not induce current. Also, there is no current density induced by a twisted scalar field, as noted from Eqs. \eqref{eq19} and \eqref{eq16}. Regarding the VEV's of the field squared and energy-momentum tensor, in the twisted scalar field case, our results are consistent with previous ones found, for instance, in Refs. \cite{PhysRevD.21.949, smith}. 

Therefore, by considering a scenario in which a charged scalar field modes propagate in a conical spacetime and are subjected to a non-periodic condition, we 
have generalized the results reported in Refs. \cite{PhysRevD.21.949, smith} where it was analysed the massless and twisted scalar field. Our results for the VEV's of the field squared and energy-momentum tensor, on the other hand, are obtained in the massive case of a charged scalar field and depend on the phase $\beta$, which includes all possible values (not only the twisted one). Furthermore, we have also obtained an induced current density which in the context of the proposed scenario is a new one. Such a current density is induced as a consequence of the symmetry obeyed by the action \eqref{action} under the condition \eqref{eq1}, in accordance with the Noether's theorem. Moreover, one should point out again that the physics associated with the quasi-periodic condition, used in the proposed scenario here, has a connection with nanotube physics. That is, the way a carbon sheet is cut can be associated with the different values of the phase angle in the interval $0\leq\beta\leq 1$.
%

{\acknowledgments} We would like to thank Eug\^enio R. Bezerra de Mello and Eduardo A. F. Bragan\c{c}a for comments on a draft of this article. K.E.L.F would like to thank the Brazilian agency CAPES for financial support. H.F.S.M acknowledges partial support from the Brazilian agency CNPq (Grants No. 305379/2017-8 and 430002/2018-1).
\bibliographystyle{ieeetr}

\begin{thebibliography}{10}

\bibitem{Milonni:1994xx}
P.~W. Milonni, {\em {The Quantum vacuum: An Introduction to quantum
  electrodynamics}}.
\newblock Boston, USA: Academic (1994) 522 p, 1994.

\bibitem{birrell1984quantum}
N.~D. Birrell and P.~C.~W. Davies, {\em Quantum fields in curved space}.
\newblock Cambridge university press, 1984.

\bibitem{bordag2009advances}
M.~Bordag, G.~L. Klimchitskaya, U.~Mohideen, and V.~M. Mostepanenko, {\em
  Advances in the Casimir effect}, vol.~145.
\newblock OUP Oxford, 2009.

\bibitem{Milton:2001yy}
K.~A. Milton, ``{The Casimir effect: Physical manifestations of zero-point
  energy},'' {\em River Edge, USA: World Scientific (2001) 301 p}, 2001.

\bibitem{Mota:2017slg}
H.~F. Mota, E.~R. Bezerra~de Mello, and K.~Bakke, ``{Scalar Casimir effect in a
  high-dimensional cosmic dispiration spacetime},'' {\em Int. J. Mod. Phys.},
  vol.~D27, no.~12, p.~1850107, 2018.

\bibitem{saharian2010topological}
A.~Saharian, ``Topological casimir effect in nanotubes and nanoloops,'' in {\em
  Quantum Field Theory Under The Influence Of External Conditions (Qfext09)
  Devoted to the Centenary of HBG Casimir}, pp.~470--474, World Scientific,
  2010.

\bibitem{Ford:1994cr}
L.~H. Ford, ``{Gravitons and light cone fluctuations},'' {\em Phys. Rev.},
  vol.~D51, pp.~1692--1700, 1995.

\bibitem{Mota:2016mhe}
H.~F. Mota, E.~R. Bezerra~de Mello, C.~H.~G. Bessa, and V.~B. Bezerra,
  ``{Light-Cone Fluctuations in the Cosmic String Spacetime},'' {\em Phys.
  Rev.}, vol.~D94, p.~024039, 2016.

\bibitem{Ford:2005rs}
L.~H. Ford, ``{Stochastic spacetime and Brownian motion of test particles},''
  {\em Int. J. Theor. Phys.}, vol.~44, pp.~1753--1768, 2005.

\bibitem{Bessa:2008pr}
C.~H.~G. Bessa, V.~B. Bezerra, and L.~H. Ford, ``{Brownian Motion in
  Robertson-Walker Space-Times from electromagnetic Vacuum Fluctuations},''
  {\em J. Math. Phys.}, vol.~50, p.~062501, 2009.

\bibitem{Gour:1998my}
G.~Gour and L.~Sriramkumar, ``{Will small particles exhibit Brownian motion in
  the quantum vacuum?},'' {\em Found. Phys.}, vol.~29, pp.~1917--1949, 1999.

\bibitem{deMello:2014ksa}
E.~R. Bezerra~de Mello, V.~B. Bezerra, A.~A. Saharian, and H.~H. Harutyunyan,
  ``{Vacuum currents induced by a magnetic flux around a cosmic string with
  finite core},'' {\em Phys. Rev.}, vol.~D91, no.~6, p.~064034, 2015.

\bibitem{Braganca:2014qma}
E.~A.~F. Bragança, H.~F. Santana~Mota, and E.~R. Bezerra~de Mello, ``{Induced
  vacuum bosonic current by magnetic flux in a higher dimensional compactified
  cosmic string spacetime},'' {\em Int. J. Mod. Phys.}, vol.~D24, no.~07,
  p.~1550055, 2015.

\bibitem{Santos:2018ttf}
W.~Oliveira~dos Santos, H.~F. Mota, and E.~R. Bezerra~de Mello, ``{Induced
  current in high-dimensional AdS spacetime in the presence of a cosmic string
  and a compactified extra dimension},'' {\em Phys. Rev.}, vol.~D99, no.~4,
  p.~045005, 2019.

\bibitem{BezerradeMello:2012nq}
E.~R. Bezerra~de Mello and A.~A. Saharian, ``{Scalar self-energy for a charged
  particle in global monopole spacetime with a spherical boundary},'' {\em
  Class. Quant. Grav.}, vol.~29, p.~135007, 2012.

\bibitem{GL}
M.~Guimaraes and B.~Linet, ``{Selfinteraction and quantum effects near a point
  mass in three-dimensional gravitation},'' {\em Class.Quant.Grav.}, vol.~10,
  pp.~1665--1680, 1993.

\bibitem{PhysRevD.65.085013}
V.~A. De~Lorenci and E.~S. Moreira, ``Classical self-forces in a space with a
  topological defect,'' {\em Phys. Rev. D}, vol.~65, p.~085013, Mar 2002.

\bibitem{Mostepanenko:1997sw}
V.~M. Mostepanenko and N.~N. Trunov, ``{The Casimir effect and its
  applications},'' {\em Oxford, UK: Clarendon (1997) 199 p}, 1997.

\bibitem{Sparnaay:1958wg}
M.~J. Sparnaay, ``{Measurements of attractive forces between flat plates},''
  {\em Physica}, vol.~24, pp.~751--764, 1958.

\bibitem{Lamoreaux:1996wh}
S.~K. Lamoreaux, ``{Demonstration of the Casimir force in the 0.6 to 6
  micrometers range},'' {\em Phys. Rev. Lett.}, vol.~78, pp.~5--8, 1997.
\newblock [Erratum: Phys. Rev. Lett.81,5475(1998)].

\bibitem{PhysRevLett.81.5475}
S.~K. Lamoreaux, ``Erratum: Demonstration of the casimir force in the 0.6 to 6
  $\mathit{\ensuremath{\mu}}m$ range [phys. rev. lett. 78, 5 (1997)],'' {\em
  Phys. Rev. Lett.}, vol.~81, pp.~5475--5476, Dec 1998.

\bibitem{mohideen1998precision}
U.~Mohideen and A.~Roy, ``Precision measurement of the casimir force from 0.1
  to 0.9 $\mu$ m,'' {\em Physical Review Letters}, vol.~81, no.~21,
  pp.~4549--4552, 1998.

\bibitem{Bressi:2002fr}
G.~Bressi, G.~Carugno, R.~Onofrio, and G.~Ruoso, ``{Measurement of the Casimir
  force between parallel metallic surfaces},'' {\em Phys. Rev. Lett.}, vol.~88,
  p.~041804, 2002.

\bibitem{MOSTEPANENKO2000}
V.~M. Mostepanenko, ``{New experimental results on the Casimir effect},'' {\em
  {Brazilian Journal of Physics}}, vol.~30, pp.~309 -- 315, 06 2000.

\bibitem{Casimir:1948dh}
H.~B.~G. Casimir, ``{On the Attraction Between Two Perfectly Conducting
  Plates},'' {\em Indag. Math.}, vol.~10, pp.~261--263, 1948.
\newblock [Kon. Ned. Akad. Wetensch. Proc.100N3-4,61(1997)].

\bibitem{BezerradeMello:2011nv}
E.~R. Bezerra~de Mello and A.~A. Saharian, ``{Topological Casimir effect in
  compactified cosmic string spacetime},'' {\em Class. Quant. Grav.}, vol.~29,
  p.~035006, 2012.

\bibitem{BezerradeMello:2011sm}
E.~R. Bezerra~de Mello and A.~A. Saharian, ``{Vacuum polarization by a flat
  boundary in cosmic string spacetime},'' {\em Class. Quant. Grav.}, vol.~28,
  p.~145008, 2011.

\bibitem{Christensen:1976vb}
S.~M. Christensen, ``{Vacuum Expectation Value of the Stress Tensor in an
  Arbitrary Curved Background: The Covariant Point Separation Method},'' {\em
  Phys. Rev.}, vol.~D14, pp.~2490--2501, 1976.

\bibitem{Wald:1978pj}
R.~M. Wald, ``{Trace Anomaly of a Conformally Invariant Quantum Field in Curved
  Space-Time},'' {\em Phys. Rev.}, vol.~D17, pp.~1477--1484, 1978.

\bibitem{smith}
A.~G. Smith, ``{On the evolution of cosmic strings},'' in {\em {The formation
  and evolution of cosmic strings : proceedings of a workshop supported by the
  SERC and held in Cambridge, 3-7 July, 1989}} (G.~W. Gibbons, S.~W. Hawking,
  and T.~Vachaspati, eds.), (Cambridge), pp.~263--292, Cambridge University
  Press, 1990.

\bibitem{Braganca:2019mvj}
E.~A.~F. Bragança, H.~F. Santana~Mota, and E.~R. Bezerra~de Mello, ``{Vacuum
  expectation value of the energy-momentum tensor in a higher dimensional
  compactified cosmic string spacetime},'' {\em Eur. Phys. J. Plus}, vol.~134,
  no.~8, p.~400, 2019.

\bibitem{Santos:2019ais}
W.~Oliveira~dos Santos, E.~R. Bezerra~de Mello, and H.~F. Mota, ``{Vacuum
  polarization in high-dimensional AdS spacetime in the presence of a cosmic
  string and compactified extra dimension},'' 2019.

\bibitem{Sriramkumar:2000nq}
L.~Sriramkumar, ``{Fluctuations in the current and energy densities around a
  magnetic flux carrying cosmic string},'' {\em Class. Quant. Grav.}, vol.~18,
  pp.~1015--1025, 2001.

\bibitem{VS}
A.~Vilenkin and E.~P.~S. Shellard, {\em {Cosmic strings and other topological
  defects}}.
\newblock Cambridge monographs on mathematical physics, Cambridge: Cambridge
  Univ. Press, 1994.

\bibitem{hindmarsh}
M.~Hindmarsh and T.~Kibble, ``{Cosmic strings},'' {\em Rept.Prog.Phys.},
  vol.~58, pp.~477--562, 1995.

\bibitem{escidoc:153364}
B.~Allen and E.~P.~S. Shellard, ``{On the evolution of cosmic strings},'' in
  {\em {The formation and evolution of cosmic strings : proceedings of a
  workshop supported by the SERC and held in Cambridge, 3-7 July, 1989}} (G.~W.
  Gibbons, S.~W. Hawking, and T.~Vachaspati, eds.), (Cambridge), pp.~421--448,
  Cambridge University Press, 1990.

\bibitem{Copeland:2011dx}
E.~J. Copeland, L.~Pogosian, and T.~Vachaspati, ``{Seeking String Theory in the
  Cosmos},'' {\em Class.Quant.Grav.}, vol.~28, p.~204009, 2011.

\bibitem{Hindmarsh:2011qj}
M.~Hindmarsh, ``{Signals of Inflationary Models with Cosmic Strings},'' {\em
  Prog.Theor.Phys.Suppl.}, vol.~190, pp.~197--228, 2011.

\bibitem{Mota:2014uka}
H.~F. Santana~Mota and M.~Hindmarsh, ``{Big-Bang Nucleosynthesis and Gamma-Ray
  Constraints on Cosmic Strings with a large Higgs condensate},'' {\em Phys.
  Rev.}, vol.~D91, no.~4, p.~043001, 2015.

\bibitem{Katanaev:1992kh}
M.~O. Katanaev and I.~V. Volovich, ``{Theory of defects in solids and
  three-dimensional gravity},'' {\em Annals of Phys.}, vol.~216, pp.~1--28,
  1992.

\bibitem{o2006carbon}
M.~J. O\textsc{\char13}connell, {\em Carbon nanotubes: properties and
  applications}.
\newblock CRC press, 2006.

\bibitem{malic2013graphene}
E.~Malic and A.~Knorr, {\em Graphene and Carbon Nanotubes: Ultrafast Optics and
  Relaxation Dynamics}.
\newblock John Wiley \& Sons, 2013.

\bibitem{phan2017functional}
N.~D.~D. Phan {\em et~al.}, ``Functional integral method in quantum field
  theory of plasmons in graphene,'' {\em Advances in Natural Sciences:
  Nanoscience and Nanotechnology}, vol.~8, no.~4, p.~045017, 2017.

\bibitem{romanovsky2013topological}
I.~Romanovsky, C.~Yannouleas, and U.~Landman, ``Topological effects and
  particle physics analogies beyond the massless dirac-weyl fermion in graphene
  nanorings,'' {\em Physical Review B}, vol.~87, no.~16, p.~165431, 2013.

\bibitem{Spinally:2000ii}
J.~Spinally, E.~R. Bezerra~de Mello, and V.~B. Bezerra, ``{Relativistic quantum
  scattering on a cone},'' {\em Class. Quant. Grav.}, vol.~18, pp.~1555--1566,
  2001.

\bibitem{PhysRevD.21.949}
L.~H. Ford, ``Twisted scalar and spinor strings in minkowski spacetime,'' {\em
  Phys. Rev. D}, vol.~21, pp.~949--957, Feb 1980.

\end{thebibliography}

\end{document}